\documentclass[twocolumn,twoside,slac_two]{revtex4}
\usepackage{graphicx}
\usepackage{fancyhdr}
\begin{document}

\title{Evidences for a double component in GRB 101023}

%

\author{A. V. Penacchioni$^{1,2}$; R. Ruffini$^{1,2,3}$; L. Izzo$^{1,3}$, M. Muccino$^{1,3}$, B. Patricelli$^{1,2}$; C. L. Bianco$^{1,3}$, L.Caito$^{1,3}$}
\affiliation{$^{1}$ Dip. di Fisica, Sapienza Universit\`a di Roma and ICRA, Piazzale
Aldo Moro 5, I-00185 Roma, Italy, \\
\null\hspace{0.5cm} E-mail: [ana.penacchioni;ruffini;luca.izzo;marco.muccino;bianco;letizia.caito]@icra.it \\
$^{2}$ Universit\'e de Nice Sophia Antipolis, Nice, CEDEX 2, Grand Chateau Parc Valrose, \\
$^{3}$ ICRANet, Piazzale della Repubblica 10, I-65122 Pescara, Italy. E-mail: barbara.patricelli@icranet.org}




\begin{abstract}
We present the results of the analysis of GRB 101023 in the fireshell scenario. Its redshift has not been determined due to the lack of data in the optical band, so we tried to infer it from the Amati Relation, obtaining z=0.9. Its light curve presents a double emission, which makes it very similar to the already studied GRB 091018. We performed a time-resolved spectral analysis with XSPEC using different spectral models, and fitted the light curve with the numerical code GRBsim. We used Fermi GBM data to build the light curve, in particular the second INa detector, in the range (8-440 keV).  We found that the first emission does not match the requirements for a GRB, while the second part perfectly agrees with being a canonical GRB, with a P-GRB lasting 4s. 
\end{abstract}

\maketitle

\thispagestyle{fancy}


\section{INTRODUCTION}

GRB 101023 has been observed by many satellites, as Konus Wind, Swift BAT and Fermi in gamma rays, Swift XRT in X-rays and Swift UVOT, Gemini and GROND in the optical band. Its redshift cannot be determined due to the lack of data, but we tried to infer it by using different methods. The source appears to be morphologically very similar to GRB 090618, with a total energy of $E_{iso} = 2.7 \times 10^{53}$ erg. The aim is to compare and contrast these two sources; to identify from this comparison the redshift of GRB 101023, and to consequently determine all the physical parameters. In this work we present the method we used to constrain the redshift. Then we proceed to examine episode 1 and episode 2 of this GRB, building its light curve and spectrum. We identify episode 2 with a canonical GRB, identifying also the P-GRB. We simulate the light curve and spectrum of this second episode with a numerical code called GRBsim. Afterwards, we go into further detail in the analysis of the first episode, making clear the evolution of the thermal component and the radius of the outermost shell, establishing the complete correspondence with GRB 090618. Finally, we present the conclusions.

\subsection{Fireshell Scenario}

In the fireshell scenario, the GRB emission originates from a process of vacuum polarization, resulting in pair creation in the so-called dyadosphere. In the process of gravitational collapse to a black hole, there forms an $e^{\pm}$ plasma in thermal equilibrium, with total energy $E_{tot}^{e^{\pm}}$ . The annihilation of these pairs occurs gradually and is confined in a shell, called "fireshellÓ. This shell self-accelerates to relativistic velocities, engulfing the baryonic matter (of mass $M_B$) left over in the process of collapsing and reaching a thermal equilibrium with it. The baryon load is measured by the dimensionless parameter $B = M_Bc^2/E_{e^{\pm}}$ . The fireshell continues to self-accelerate up to relativistic velocities until it reaches the transparency condition. At this time we have a first flash of radiation, the Proper-GRB (P-GRB). The energy released in the P-GRB is a fraction of the initial energy of the dyadosphere $E_{dya}$. The residual plasma of leptons and baryons interacts with the circumburst medium (CBM) as it expands, giving rise to a multi-wavelength emission, the extended afterglow. The structures observed in the prompt emission of a GRB are due to the inhomogeneities in this CBM. In this way we need few parameters for a complete description of a GRB: the dyadosphere energy $E_{dya}$, the baryon load $B$ and the CBM density distribution, $n_{CBM}$.

\section{DATA ANALYSIS}

\subsection{Data reduction}

To obtain the Fermi GBM light-curve and spectra in the band (8 - 440) KeV (see Fig. \ref{lightcurve}) we made use of the standard headas procedure. We downloaded the data from the gsfc website\footnote{http://heasarc.gsfc.nasa.gov/heasoft} and constructed the count light curve with gtbindef and gtbin tools, to obtain a GTI file for the energy distribution and the light curve, respectively. We fixed the time binning in 1s. We used the data from the second INa detector in the range 8- 440 KeV. Once the light curve was created, we subtracted the background and made the starting time coincide with the trigger time, of 309567006.72 in MET seconds. After the reduction of data was complete, we proceeded with its time resolved spectral analysis with the program XSPEC.

\subsection{light curve}
\begin{figure}
\centering
\includegraphics[width=0.95\hsize]{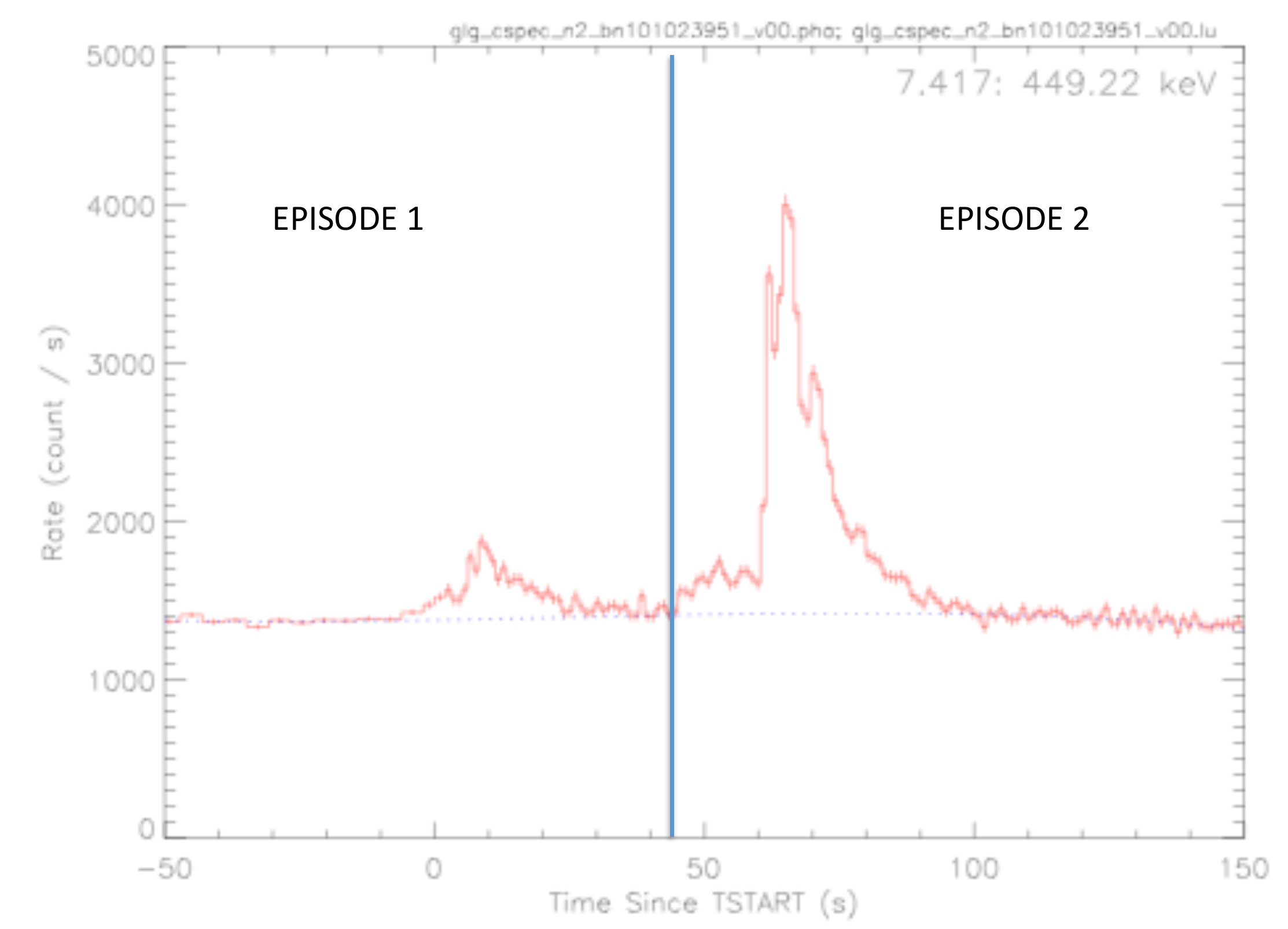}
\caption{Count light curve of GRB 101023 obtained from Fermi GBM detector, with a bin time of 1 s. The time is given with respect to the GBM trigger time of 22:50:04.73 UT, 23 October, 2010. The plot was obtained with the RMFIT program. The two-episode nature of the GRB is evidenced in analogy with GRB 090618.}
\label{lightcurve}
\end{figure}

The GBM light curve (Fig. \ref{lightcurve}) shows two major pulses. The first one starts at the trigger time and lasts 25 s. It consists of a small peak that lasts about 10s, followed by a higher emission that decays slowly with time. The duration, as well as the topology of this curve, lead to think that this may not be the case of a canonical GRB, but its origin may lie on another kind of source, which remains still unidentified. The second pulse starts at 45s after the trigger time and lasts 44s. It presents a peaky structure, composed by a short and weak peak at the beginning, followed by several bumps, big not only in magnitude but also in duration. This second emission, on the contrary, does have all the characteristics which describe a canonical GRB.

\section{REDSHIFT DETERMINATION}

The redshift of this source is unknown, due to the lack of data in the optical band. However, we can infer a value of z=0.9 under the hypothesis that, as every long GRB, it should follow the Amati relation \citep{Amati}. This relates the isotropic energy $E_{iso}$ emitted by a GRB to the peak energy in the rest frame $E_{p,i}$ of its $\nu F \nu$ electromagnetic spectrum. $E_{iso}$ is the isotropic equivalent radiated energy, while $E_{p,i}$ is the photon energy at which the time averaged $\nu F \nu$ spectrum peaks. $E_{iso}$ is given by:

\begin{equation}
\label{Eiso}
E_{iso}=\frac{4 \pi d_l^2}{(1+z)} S_{bol},
\end{equation}
where $d_l^2$ is the luminosity distance, $z$ is the redshift and $S_{bol}$ is the bolometric fluence (estimated between $1$ keV and $10$ MeV), related to the observed fluence in a given detection band ($E_{min}$, $E_{max}$) by
\begin{equation}
S_{bol}=S_{obs}\frac{\int^{10 MeV / (1+z)}_{1 keV/ (1+z)} E \phi(E) dE}{\int ^{E^{max}}_{E^{min}} E \phi(E) dE},
\end{equation}
with $\phi$ the spectral model considered for the spectral data fit. $E_{p,i}$ is related to the peak energy $E_p$ in the observer frame by
\begin{equation}
E_{p,i}=E_p (1+z)
\end{equation}

\begin{figure}
\centering
\includegraphics[width=0.95\hsize]{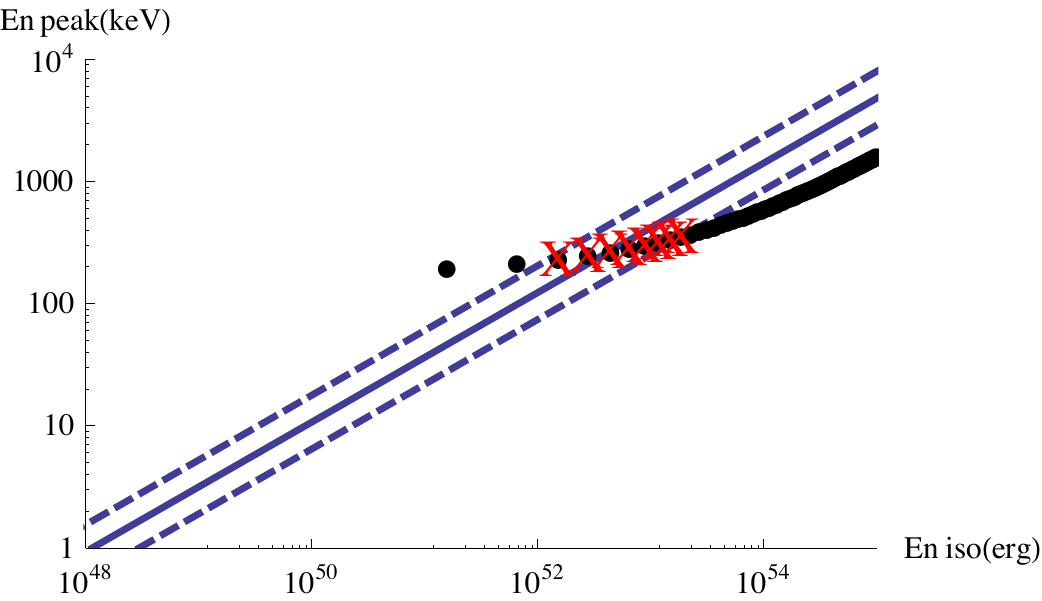}
\caption{Plot of the relation between $E_{p,i}$ and $E_{iso}$ for the second episode of GRB 101023, considering different values of the redshift. It can be seen that the plot lies within $1\sigma$ for the range z= 0.3 - z= 1.0.}
\label{PICTURE OF AMATI RELATION}
\end{figure}

We started our analysis under the hypothesis that episode 2 is a long GRB. We computed the values of $E_{p,i}$ and $E_{iso}$ for different given values of $z$ and plotted them in Fig \ref{PICTURE OF AMATI RELATION}. We found that the Amati relation is fulfilled by episode 2 for $0.3 < z < 1.0$. This interval has been calculated at $1 \sigma$ from the best fit from the Amati relation. We chose z = 0.9 because for this value of the redshift the energetics of this source are very similar to those of GRB 090618, whose redshift is known to be z=0.54. This is, the isotropic energy of events 1 and 2 in GRB 090618 are exactly the same as the isotropic energy of GRB1 and GRB2 in GRB 101023, respectively.

\section{SPECTRAL ANALYSIS}

To proceed with the fitting of the spectra, we defined first of all the time intervals we wanted to analyze making use of the gtbindef tool. We obtained a GTI file for the energy distribution. Secondly, we used gtbin to obtain the spectrum and in that way be able to fit the data with different models, via XSPEC. We used a Black body plus a power-law model and a Band model. The results are shown in Table \ref{TABLA}.

 \begin{table*}
\centering
\caption{Time-resolved spectral analysis of GRB 101023. We have considered two main time intervals, corresponding to episode 1 and episode 2 in the light curve. We have proposed for each time interval a Band spectral model \citep{Band} and a Black body plus a power-law model.} 
\label{TABLA} 
\begin{tabular}{l c c c c c| c c c c c }
\hline\hline
Time $[s]$& $\alpha$        & $\beta$                & $E_0^{BAND}[\rm{keV}]$       & $\chi^2$ & Norm                   &$kT [\rm{keV}]$         &$\gamma$      &$\chi^2$& Norm $^{po}$ & Norm $^{BB}$\\
0-44                      & -1.3$\pm$0.8& -1.9$\pm$0.2       & 87$\pm$147           & 0.98        & 0.006$\pm$0.01&14$\pm$6 & -1.7$\pm$0.1    &0.98      &0.0003$\pm$0.0004 & (4.1 $\pm7.4) \times 10^{-5}$\\
45-89                    &-0.9$\pm$0.1 & -2.02$\pm$0.1     &151$\pm$24            & 1.09        &0.043$\pm$0.008&26$\pm$1&-1.58$\pm$0.03&1.12   &0.0124$\pm$0.0006& (4.2 $\pm1.1)\times 10^{-5}$\\
\hline
\hline
\end{tabular}
\end{table*}

\begin{figure}
\centering
\includegraphics[width=0.95\hsize]{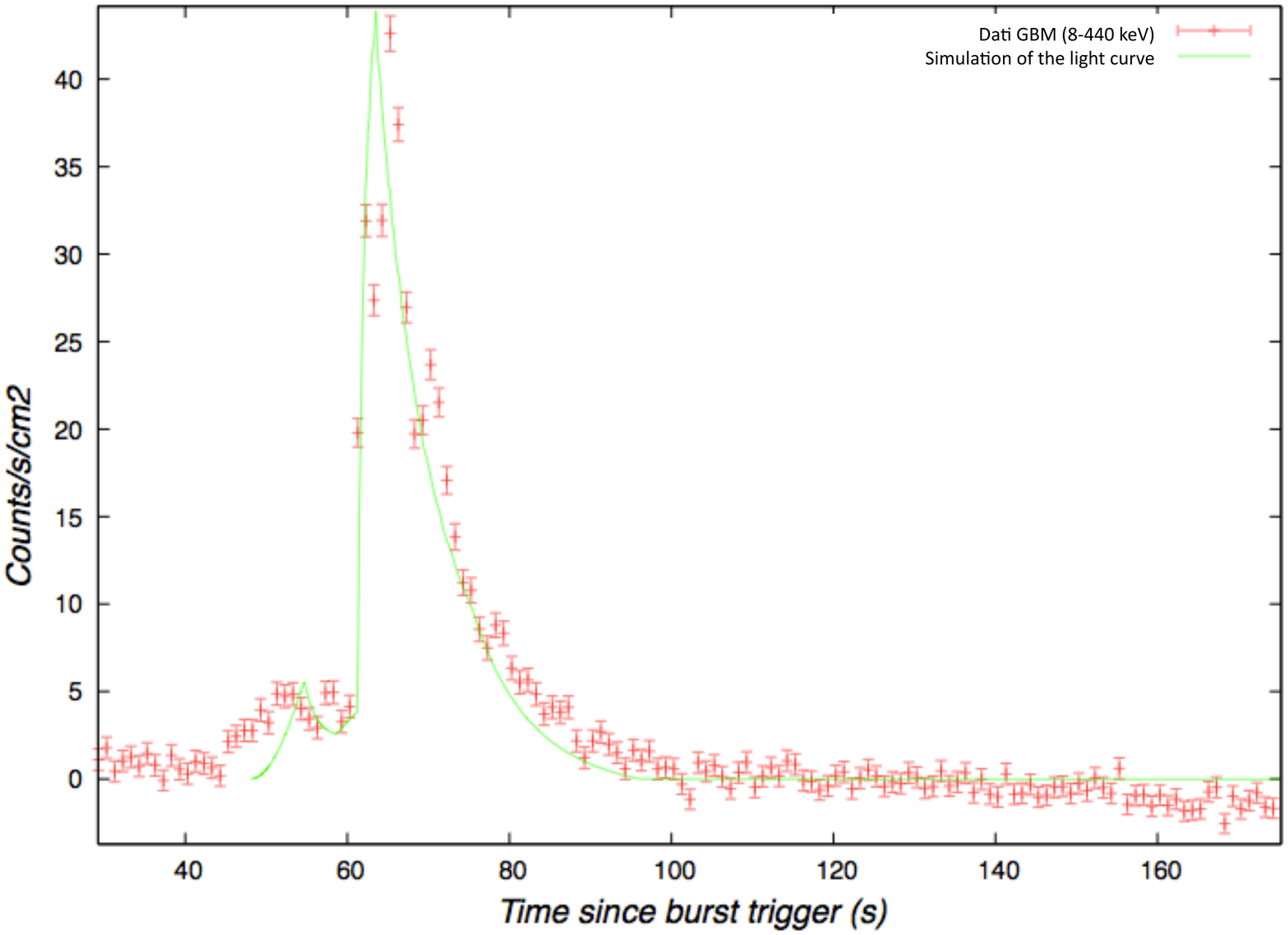}
\caption{Fit of the second major pulse of the light curve of GRB 101023.}
\label{simulation of the light curve}
\end{figure}

\begin{figure}
\centering
\includegraphics[width=0.95\hsize]{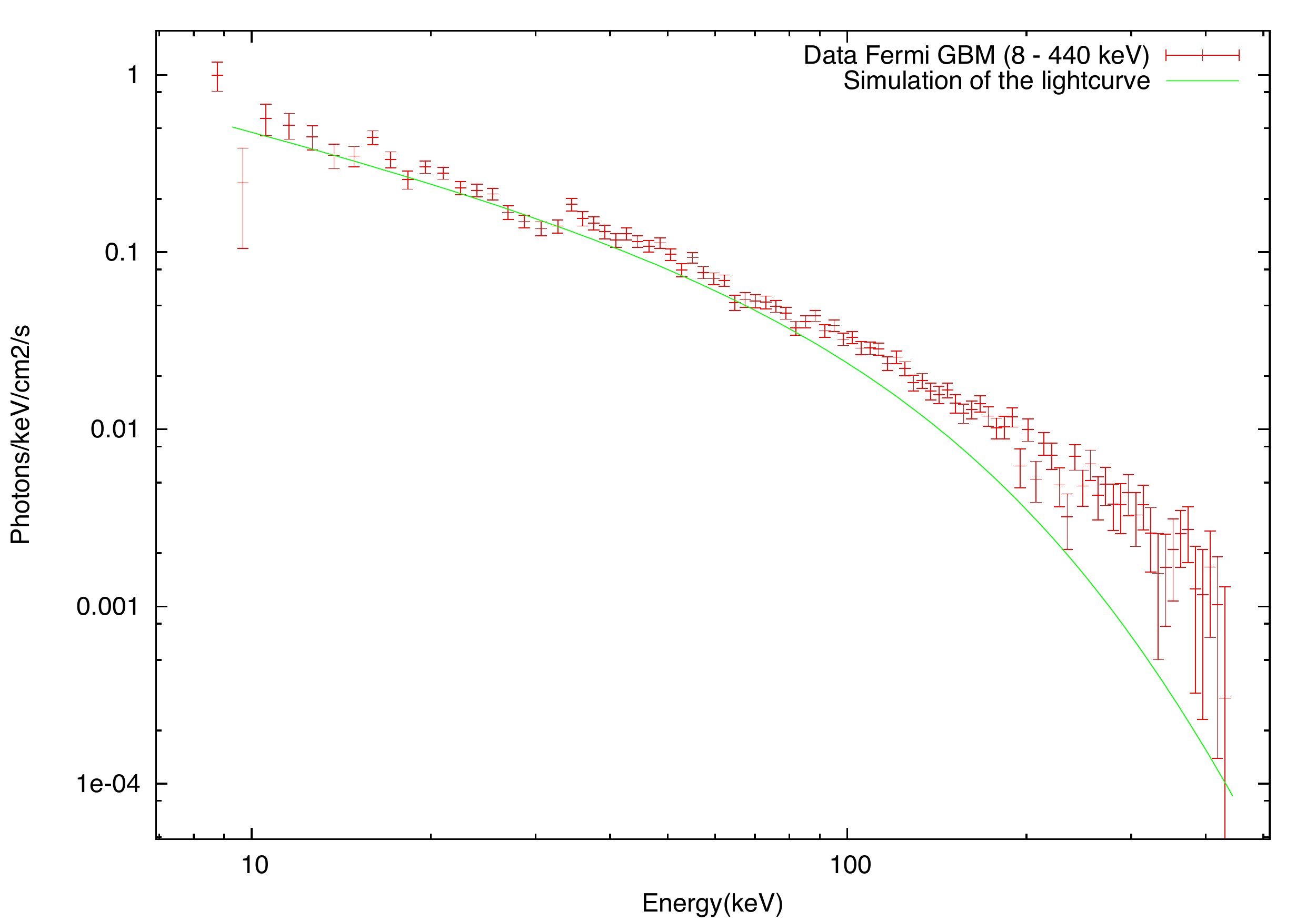}
\caption{Fit of the spectrum of Episode 2.}
\label{FITTED SPECTRUM}
\end{figure}

We can see that the second emission can be very well fitted by a Black body plus a power-law model. In particular, we find for the P-GRB an observed temperature of $kT= 13.5 \pm 2.8$ keV, a photon index of $\gamma= 2.5 \pm 0.7$ and a $\chi^2 =0.91$.  From these values we can infer a P-GRB energy $E_{P-GRB}= 1.32 \times 10^{51}$ erg and a total energy of $E_{iso}= 1.65 \times 10^{53}$ erg, i.e. the P-GRB energy release is the $0.7 \%$ of the total. We introduced these results in a numerical code to simulate the light curve (see Fig. \ref{simulation of the light curve}) and we found the following values for the parameters in the fireshell model: $B= 4.8 \times 10^{-3}$, $kT_{th} = 20.83$ keV, $\Gamma= 206.81$, and a Laboratory radius of $1.44 \times 10^{14}$ cm. We also simulated the spectrum of episode 2, shown in Fig. \ref{FITTED SPECTRUM}. 

\subsection{Episode 1: Temperature and radius evolution}

\begin{figure}
\centering
\includegraphics[width=0.95\hsize]{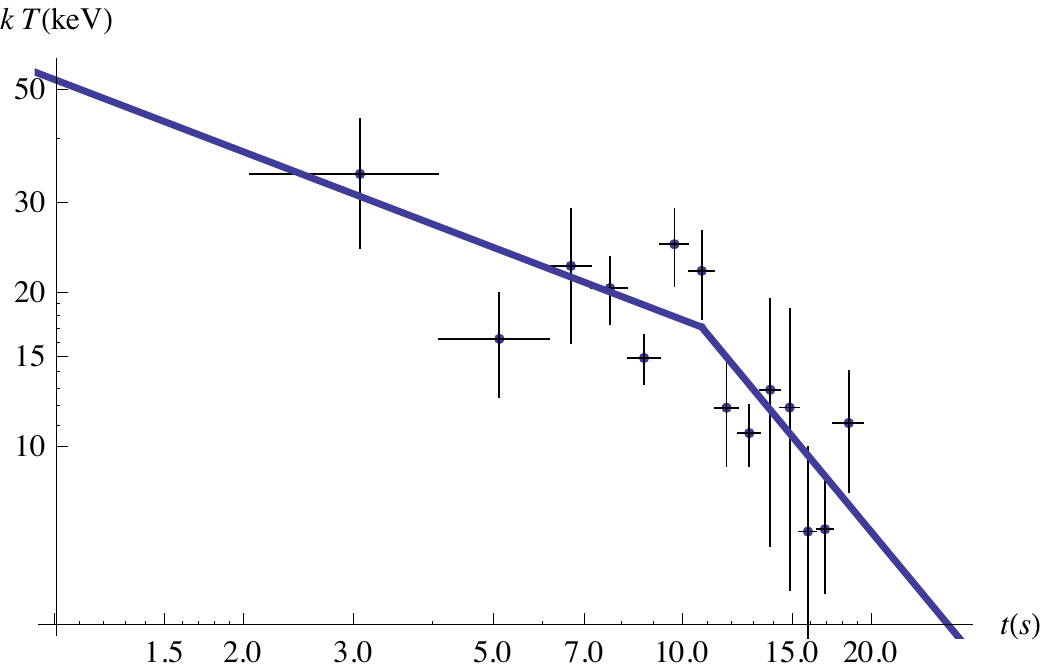}
\caption{Evolution of the observed temperature kT of the BB component. The blue line corresponds to a broken power-law fit. The indices of the first and second power-law are $\alpha=-0.47 \pm 0.34$ and $\beta= -1.48 \pm 1.13$, respectively. The break occurs at 11 s after the trigger time.}
\label{kTepisode1}
\end{figure}

\begin{figure}
\centering
\includegraphics[width=0.95\hsize]{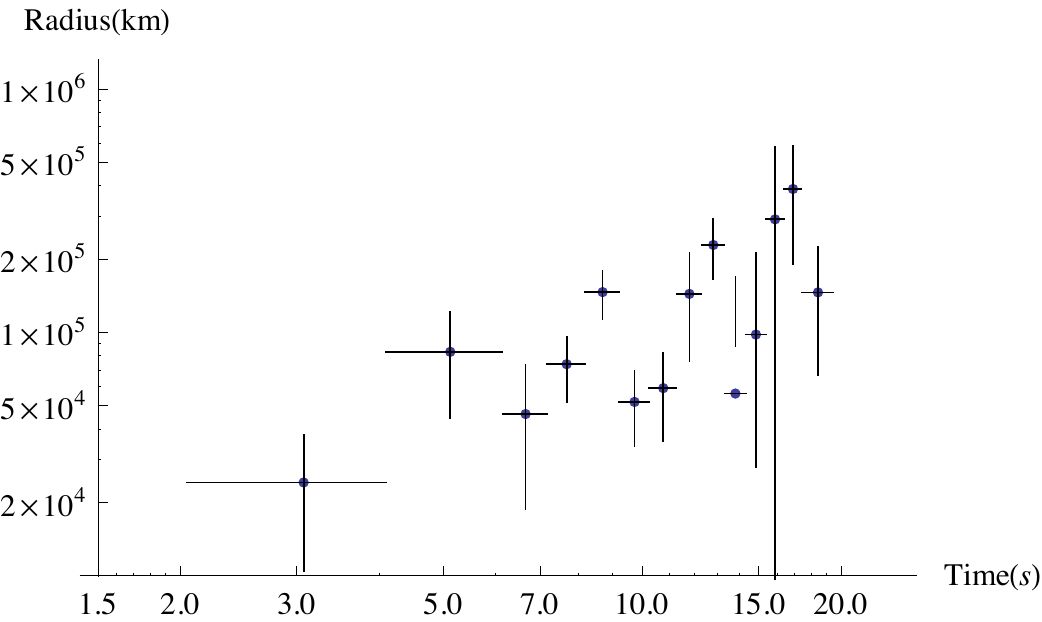}
\caption{Evolution of the radius of the first episode progenitor.}
\label{radius}
\end{figure}

To analyze episode 1 more into detail, in order to identify the nature of this phenomenon, we plotted the temperature of the Black body component as a function of time, for the first 20 s of emission (see Fig. \ref{kTepisode1}). We note a strong evolution in the first 20 s of emission which, according to \citet{Ryde2004} can be reproduced by a broken power-law behavior, with $\alpha=-0.47 \pm 0.34$ and $\beta= -1.48 \pm 1.13$ being the indices of the first and second power-law, respectively. We also plotted the radius of the most external shell with time (see Fig. \ref{radius}). Following \citet{Luca}, the radius can be written as

\begin{equation}
r_{em}=\frac{\hat{R} D}{(1+z)^2},
\end{equation}
where $\hat{\rm{R}}= \phi_{obs} /(4 \pi \sigma T_{obs}^4)$ is a parameter, $D$ is the luminosity distance, $\Gamma$ is the Lorentz factor and $\phi_{obs}$ is the observed flux. We can see that the radius remains almost constant (in fact it increases, but only slightly). From this it is possible to see that the plasma is expanding at non-relativistic velocities. According to the work of \citet{Arnett}, there is an expansion phase of the boundary layers, while the iron core suffers a contraction. This is due to the presence of strong waves originated while the different shells of the progenitor mix during the collapse phase. This fact confirms the non-GRB nature for the first episode.

\section{CONCLUSIONS}

GRB 101023 is a very interesting source for the following reasons:

1) It has been observed by many satellites in different energy bands. Despite the wide energy range coverage, there is no redshift determination for this source, which complicates the determination of its characteristic parameters. This has challenged the indirect inference of its cosmological redshift.

2)Morphologically, there is a striking similarity between GRB 101023 and GRB 090618, as can be seen from the light curves. Following the study of GRB 090618, we have divided the emission into two episodes: Episode 1, that lasts 45 s, presents a smooth emission, without spikes, that decays slowly with time. Episode 2, of 44 s of duration, presents a spiky structure, composed of a short and small peak at the beginning, followed by several intense bumps, after which there is a fast decay with time. Episode 2 has all the characteristics of a canonical long GRB.

3)Since we do not know the redshift, we cannot infer the total energy of the source, $E_{iso}$. We have attempted to infer the cosmological redshift of the source from the peak energy $E_{peak}$, using the Amati relation under the hypothesis that Episode 2 is a canonical long GRB. We constrain the value of the redshift to be between 0.3 and 1.0. We assumed for the redshift of this source  $z=0.9$. 

4)We performed a time-resolved analysis of Episode 1. We fitted a Black body plus a power law model and plotted the evolution of the black body component with time. The observed temperature decreases during the first 20 s following a broken power-law: the first with index  $\alpha=-0.47 \pm 0.34$ and the second with index $\beta= -1.48 \pm 1.13$. From the observed Black body temperature, we infer the radius of the Black body emitter and its variation with time, see Fig. \ref{kTepisode1}. We have seen that it increases slightly during the first 20 s of emission. In analogy with GRB 090618, we conclude that Episode 1 originates from the last phases of gravitational collapse of a stellar core, just prior to the collapse to a black hole. We call this core a ``proto-black hole" \citep{Ruffini2010a}. Immediately after, the collapse occurs and the GRB is emitted (Ep. 2). 

5) From the knowledge of the redshift of the source, we have analyzed Episode 2 within the fireshell model. We determined a total energy $E_{iso}=1.79 \times 10^{53}$ erg and a P-GRB energy of $2.51 \times 10^{51}$ erg, which we used to simulate the light curve and spectrum with the numerical code GRBsim. We find a baryon load $B=3.8 \times 10^{-3}$ and, at the transparency point, a value of the laboratory radius of $1.34 \times 10^{14}$ cm, a theoretically predicted temperature of $kT_{th}=13.26$ keV (after cosmological correction) and a Lorentz Gamma factor of $\Gamma=260.48$, confirming that Episode 2 is indeed a canonical GRB.

We conclude that GRB 101023 and GRB 090618 have striking analogies and are members of a specific new family of GRBs originating from a single core collapse. The existence of precise scaling laws between these two sources opens a new window on the use of GRBs as distance indicators. We are planning further identifications of additional sources belonging to this family at higher redshifts. 

Full details on the analysis of GRB 101023 can be found in Penacchioni et al., submitted to A\&A.

\subsection{Acknowledgments} \label{Ack}
We thank the Swift and Fermi team for their support. This work made use of data supplied by the UK Swift data Centre at the university of Leicester.

\bigskip 

\end{document}